\documentclass{article}
\usepackage[T1]{fontenc} % add special characters (e.g., umlaute)
\usepackage[utf8]{inputenc} % set utf-8 as default input encoding
\usepackage{kotex}
\usepackage{ismir,amsmath,cite,url}
\usepackage{graphicx}
\usepackage{color}
\usepackage{multirow}
\usepackage{caption}
\usepackage{booktabs}
\usepackage{enumitem}
\usepackage{wasysym}
\usepackage{harmony}
\usepackage{lineno}

\title{Six Dragons Fly Again: Reviving 15th-Century Korean Court Music with Transformers and Novel Encoding}
\multauthor
{Danbinaerin Han$^{1*}$
\thanks{* Work mainly done during her master's at Sogang University}
\hspace{1.5cm} Mark Gotham$^2$ \hspace{1.4cm} Dongmin Kim$^3$ 
}
{\bfseries{\hspace{0.2cm} Hannah Park$^4$ \hspace{2.2cm} Sihun Lee$^3$ \hspace{1.6cm} Dasaem Jeong$^4$}\\ $^1$ Gradaute School of Culture Technology, KAIST, Daejeon, South Korea\\
$^2$ Deptartment of Digital Humanities, King's College London, UK \\ 
$^3$ Dept. of Artificial Intelligence, $^4$ Dept. of Art \& Technology, Sogang University, Seoul, South Korea\\
\tt\small naerin71@kaist.ac.kr, mark.gotham@kcl.ac.uk, \{dmkim, hannah, sihunlee, dasaemj\}@sogang.ac.kr}

\def\authorname{D. Han, M. Gotham, D. Kim, H. Park, S. Lee, and D. Jeong}

\usepackage[bookmarks=false,pdfauthor={\authorname},pdfsubject={\papersubject},hidelinks]{hyperref}
\begin{document}

\maketitle

\begin{abstract}
% Despite the expanding research on symbolic music generation, there has  research working on non-Western music. 
We introduce a project that revives a piece of 15th-century Korean court music, \textit{Chihwapyeong} and \textit{Chwipunghyeong}, composed upon the poem \textit{Songs of the Dragon Flying to Heaven}. One of the earliest examples of \textit{Jeongganbo}, a Korean musical notation system, the remaining version only consists of a rudimentary melody. Our research team, commissioned by the National Gugak (Korean Traditional Music) Center, aimed to transform this old melody into a performable arrangement for a six-part ensemble. 
Using \textit{Jeongganbo} data acquired through bespoke optical music recognition, we trained a BERT-like masked language model and an encoder-decoder transformer model.
We also propose an encoding scheme that strictly follows the structure of \textit{Jeongganbo} and denotes note durations as positions. 
The resulting machine-transformed version of \textit{Chihwapyeong} and \textit{Chwipunghyeong} were evaluated by experts and performed by the Court Music Orchestra of National Gugak Center. Our work demonstrates that generative models can successfully be applied to traditional music with limited training data if combined with careful design.

\end{abstract}

\section{Introduction}

\label{sec:introduction}

\textit{Six dragons fly on the east land; every endeavour is a heavenly blessing}. This is the first line of lyrics in \textit{Yongbieocheonga}, the first text written in the Korean alphabet (Hangul, 한글). Sejong the Great, one of the most respected figures in Korean history, invented and introduced Hangul in 1446. In addition to this remarkable achievement, he ordered scholar-officials to write \textit{Yongbieocheonga}, and composed music to accompany the lyrics.
% written in the newly invented , aiming to promote its widespread use. 
Three other pieces composed at the time are \textit{Yeo-Min-Lak}, \textbf{\textit{Chi-Hwa-Pyeong}} and \textbf{\textit{Chwi-Pung-Hyeong}}. These compositions are still preserved in the \textit{Veritable Records of Sejong}, which is the oldest surviving musical score in Korea\cite{TheRhythmicInterpretation}.
More detailed information is available here\cite{NationalGugakCenter2024}.
% \href{https://www.gugak.go.kr/site/program/board/basicboard/view?menuid=001003002005&pagesize=10&boardtypeid=24&boardid=13154&lang=en}
% {here}.
% 
\begin{figure}[!t]

 \includegraphics[alt={The image shows the entire framework of a research process described in a paper. The process involves converting modern Korean court music scores (jeongganbo) into a machine-readable dataset using Optical Music Recognition (OMR) and token encoding. The 15th century's old melody is infilled by a BERT-like Masked Language Model (MLM) and further processed by a Transformer language model to produce melodies for six traditional Korean musical instruments. The instruments depicted are gayageum, daegeum, ajaeng, haegeum, piri and geomungo.},width=\columnwidth]{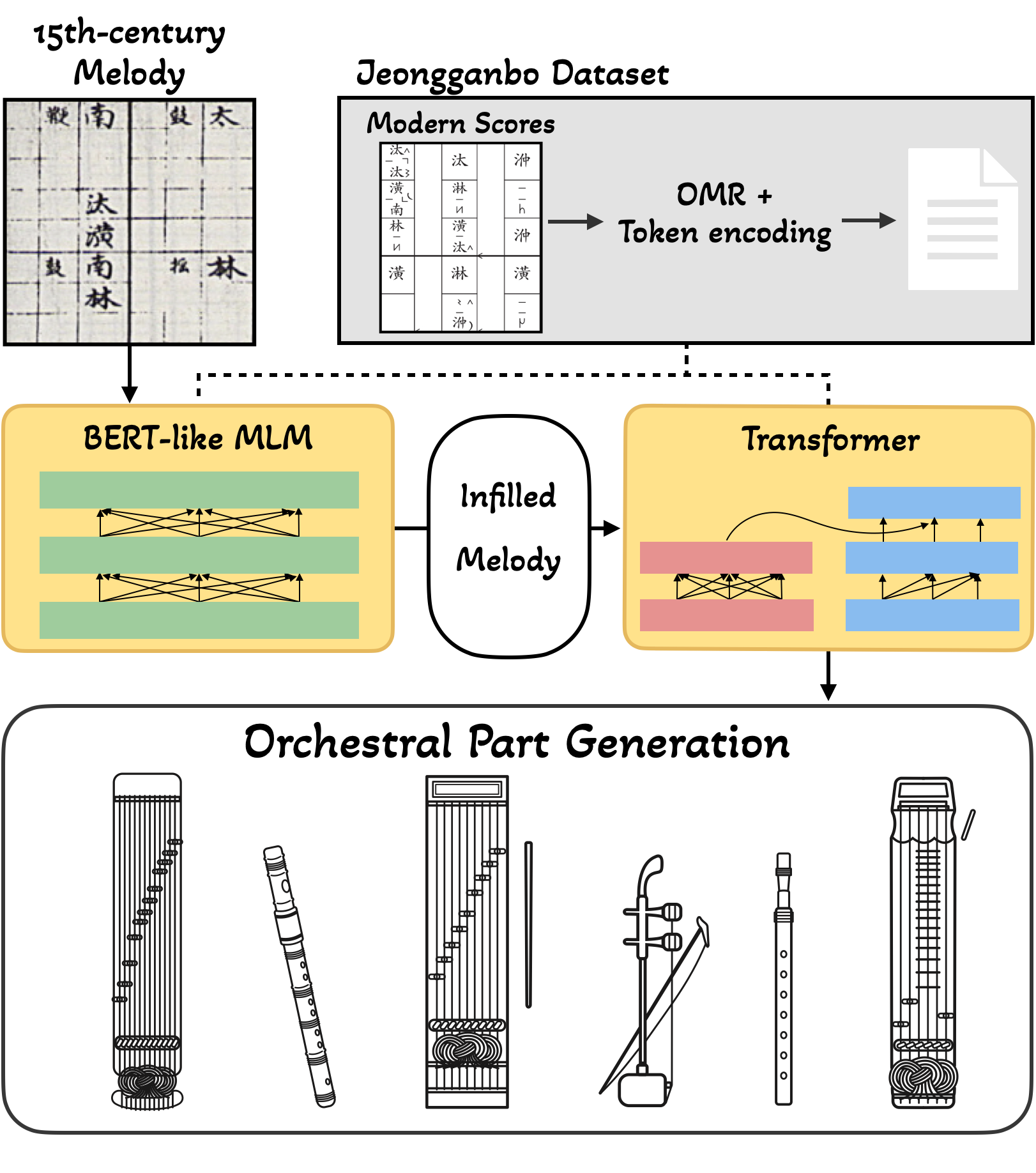}
  
  \caption{Overview of the proposed research framework}
  \label{fig:Overview}
  % \vspace{-4mm}
\end{figure}

Among these three pieces, only \textit{Yeominlak} is handed down to the present day, while the other two are no longer performed.
The National Gugak Center\footnote{\textit{Gugak} (국악) is the Korean term for traditional music}, which is the primary organization dedicated to the preservation and development of traditional music, commissioned the task of reconstructing these two pieces in a performable format using artificial intelligence systems. 
Given a simple melody of 512 \textit{gaks} (measures) of \textit{Chihwapyeong} or 132 gaks of \textit{Chwipunghyeong}, the system must generate scores for six different instruments.

Our solution encompasses a wide range of tasks in the field of music information retrieval—constructing a specialized dataset, optical music recognition, designing a domain-specific encoding scheme, training models with limited data, and generating music of concert-level quality. In this paper, we present in detail the different frameworks used in the project: two types of transformer-based models; a symbolic dataset of Korean court music acquired through optical music recognition; and a novel ``Jeonggan-like'' encoding method that notates monophonic melody by combining notes' position and pitch, along with a beat counter that informs the transformer the temporal position. 
The effectiveness of the proposed techniques was validated through quantitative metrics and subjective evaluation by experts from the National Gugak Center. Finally, we introduce a web demo that allows users to examine and generate traditional Korean court music interactively.

This project has significance not only for cultural preservation but also for wider considerations in machine learning and music generation. 
One of the many benefits to be had from the inter-cultural study of music is the different perspectives expressed in `the music itself’ as well as any notational and/or theoretical traditions that go alongside it. 
As presented in previous research \cite{micchi2020not}, the encoding of music makes significant differences in machine learning tasks.
In thinking through different ways of digitally encoding music, we stand to learn a great deal from the various syntaxes that have been used in diverse traditional contexts.

\section{Related Works}

Recent advances in neural network-based music generation have resulted in much artistic output. Since 2020, the \textit{AI Music Generation Challenge}\cite{sturm2023ai} has been held annually, focusing on generating songs in the style of Irish and Swedish folk music. This event has allowed for exploration of new methods for generation and evaluation of traditional music through the means of deep learning models.
% TODO : fill in the other examples
% Similar to the goal of our research, reviving unfinished piece with machine learning, many attempts were succesfully given 

The \textit{Beethoven X project}~\cite{gotham2022beethoven} utilized neural networks to learn Beethoven's compositional style and complete his unfinished 10th Symphony. The resulting work has been performed by an orchestra—a project outline similar to that of ours.

Attempts at automatic generation have been made for traditional music from beyond the West, including Persia \cite{ebrahimi2019procedural} and China \cite{luo2020mg}. The limited progress in such areas is often due to the distinctive traditional musical systems that demand deep understanding and unique methodologies. Such idiosyncrasies put much interest and meaning in the computational research of traditional music, since it can present new methods and perspectives to the field as a whole, while also helping preserve diverse musical heritages.

\begin{figure}[!t]
  \hspace*{-0.7cm}
   \includegraphics[alt={The image contains two corresponding musical scores. The lower part shows An example of Jeongganbo in the original notion, and the upper part displays a broadly equivalent conversion to Western classical notion. These two scores are aligned with light blue dotted lines between them.},width=1.1\columnwidth]{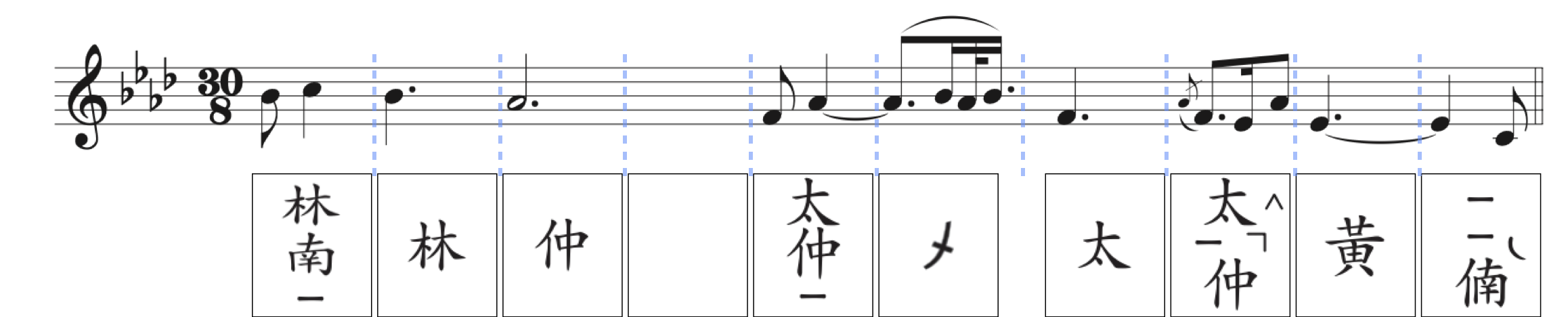}

  \caption{An example of Jeongganbo in the original notion (below) and a broadly equivalent conversion to Western classical notion (above).
  Dashed lines are part of neither notation and added simply to clarify the temporal alignment between the two systems.}
  \label{fig:jeonggan_example}
  % \vspace{-2mm}
\end{figure}

\section{Jeongganbo Dataset}

\subsection{Jeongganbo Notation}
As depicted in Figure~\ref{fig:Overview}, Korean court music is performed on a variety of instruments, including plucked string instruments (\textit{Gayageum} and \textit{Geomungo}), bowed string instruments (\textit{Haegeum} and \textit{Ajaeng}), and wind instruments (\textit{Daegeum} and \textit{Piri}), among others. These instruments are played together in a heterophonic texture, with each instrument employing its distinctive playing techniques and ornamentations. 

\begin{figure}[!t]
\centering
 \includegraphics[alt={This image shows five examples of how to interpret the rhythm based on the position of characters within jeonggan. Each position is assigned a number from 0 to 15. If there is only one character in the center of the square, it is interpreted as a dotted quarter note. The other four examples show how the rhythm is divided based on the row and column positions within the jeonggan.},width=0.8\columnwidth]{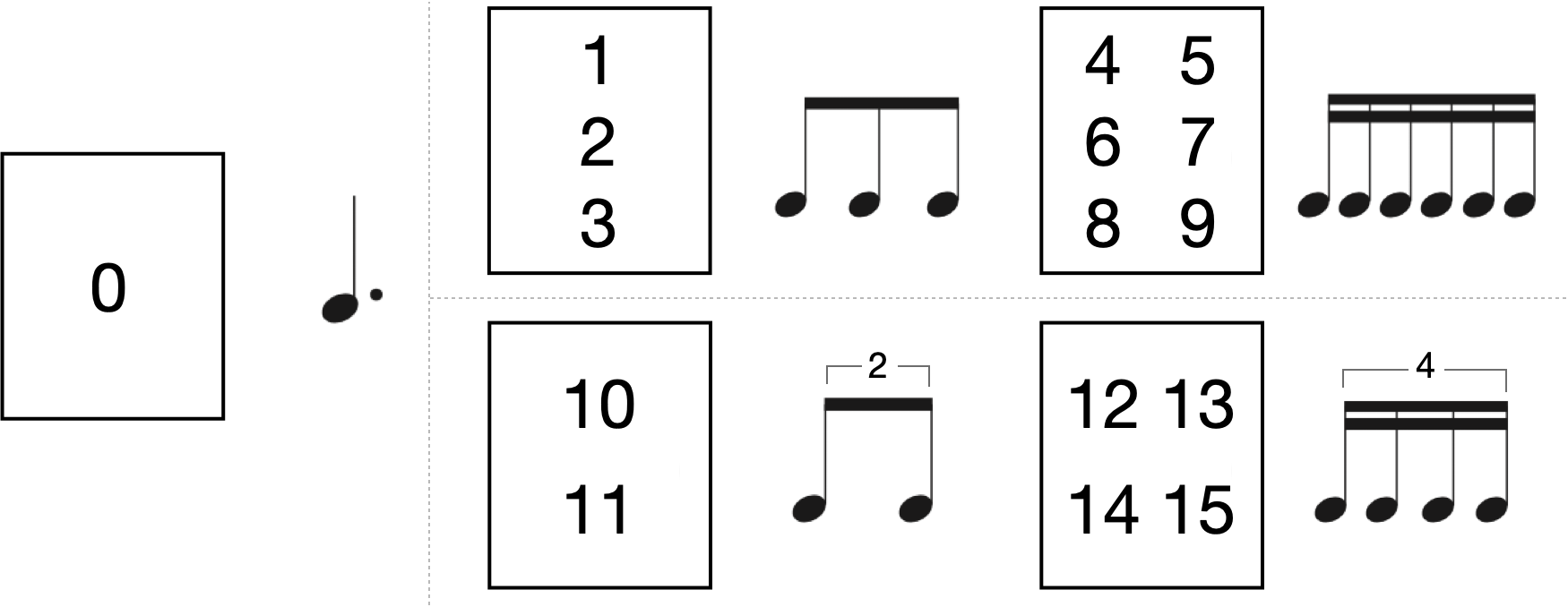}  
  \caption{Jeonggan-like encoding position labels}
  \label{fig:encoding-position}
% \vspace{-3mm}
\end{figure}

Much of Korean court music is written in \textit{\textbf{Jeongganbo}}, a traditional musical notation system. \textit{Jeongganbo} is recognized as the first system in East Asia capable of simultaneously representing both pitch and duration of notes~\cite{gnanadesikan2008writing, KoreanNotationSystem}. This versatility has been instrumental in passing down court music throughout history~\cite{koehler2015traditional}. 

% As shown in Figure~\ref{fig:jeonggan_example},
\textit{Jeongganbo} uses grid-divided boxes (\textit{Jeonggans}) as the basic unit of time. 
The \textit{number} of characters (notes) 
and their \textit{position} within each \textit{jeonggan} varies to denote rhythm.
Figure \ref{fig:jeonggan_example}
%and \ref{fig:encoding} 
provides an example passage, and figure \ref{fig:encoding-position} provides a schematic overview of possible positions.

Here, we provide a broad introduction to this rhythmic notation system in quasi-Western musical theoretic language.
Each \textit{jeonggan} is broadly equivalent to a beat.
If a \textit{jeonggan} features only one character, this note event starts at the beginning of the beat and lasts the beat's full duration.
The first box (`0') in figure \ref{fig:encoding-position} is in this form as is the second \textit{jeonggan} of figure \ref{fig:jeonggan_example} where the `compound beats' correspond to the duration $\quarternote \Pu$
(in this case for the note B$\flat$4).
At the next metrical level we have the `column' division of the `rows'.
This number of `rows' relates broadly to the top level division of the beat.
The use of three vertically stacked characters refers to 3 equal divisions of this beat (here, 3 x $\eighthnote$s).

For example, in figure \ref{fig:encoding-position}, the numbers 4–9 feature a 3-part division of the $\quarternote \Pu$ beat into 3 x $\eighthnote$s (positions 4, 6, 8), and a 2x division of those $\eighthnote$s (e.g., 4–5). If the following \textit{jeonggan} is empty, the previously played note is sustained.

Playing techniques and ornamentations called \textit{sigimsae} are sometimes notated for each instrument. When sigimsae are placed to the right of notes, they serve as ornamentations or embellishments for the corresponding note; when written on their own, they indicate timed instructions to play a specific note or musical phrase. For convenience, the example score is notated horizontally, but in practice, the score page is read from top to bottom and right to left. A line in \textit{Jeongganbo} can consist of anything from four to twenty beats, with each line representing a phrase unit.

\begin{figure}[!t]
\centering
  \includegraphics[alt={
This table compares three methods for encoding jeongganbo: 'JG-like,' 'REMI-like,' and 'ABC-like.'},width=0.9\columnwidth]{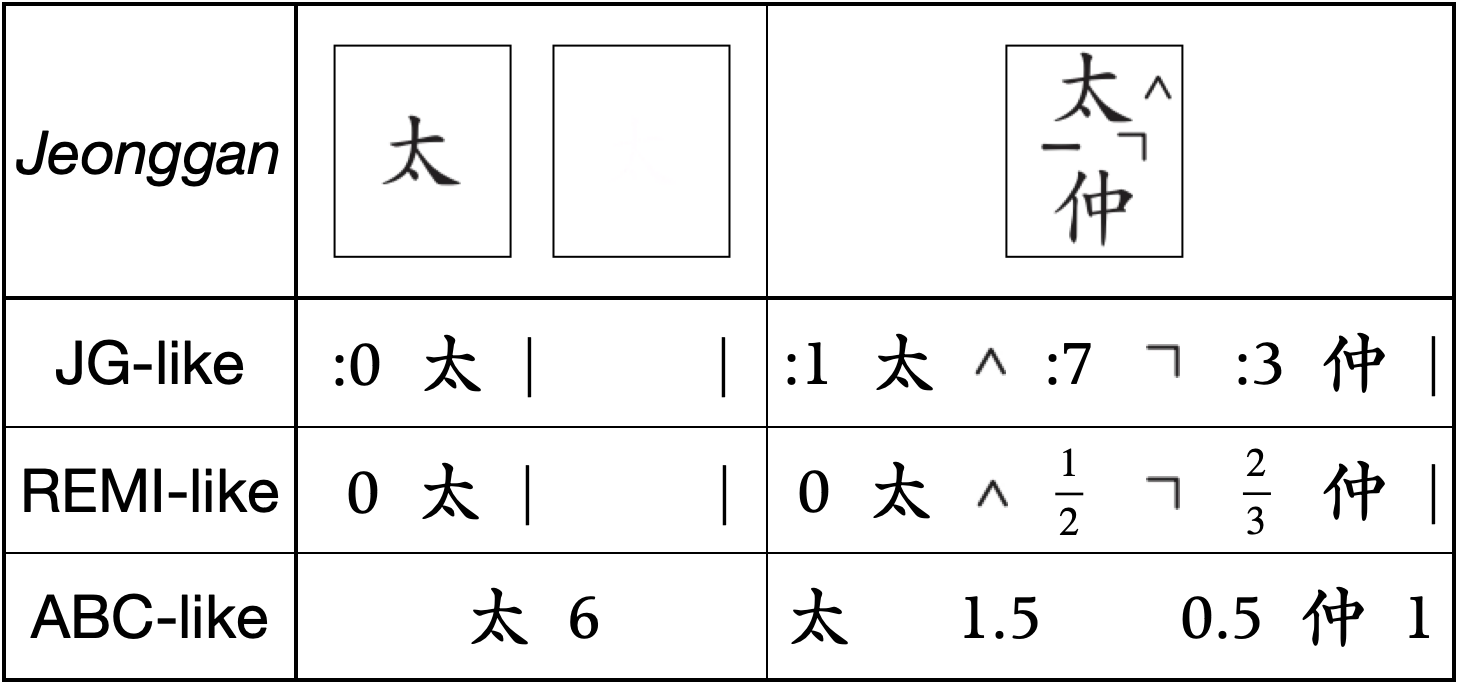}

  \caption{Comparison between encoding schemes}
  \label{fig:encoding}
  % \vspace{-3mm}
\end{figure}

\subsection{Machine Readable Dataset}
We have constructed a dataset of 85 pieces by applying optical musical recognition (OMR) to all compositions available within the manuscripts published by the National Gugak Center. The manuscripts cover the entire repertoire of remaining Korean court music\footnote{The term ``court music'' used in this paper originally refers to \textit{Jeong-ak}. Jeong-ak includes not only court music but also salon music and military music. 
However, for readability, we use ``court music'' here.}. 
OMR was necessary since the scores are only provided as PDF images and the semantic data is unavailable. 
We implemented and trained an encoder-decoder transformer with CNN 
% by synthesizing various \textit{Jeonggan} images through a rule-based approach~\cite{kim2024automatic}.
by synthesizing various \textit{Jeonggan} images in a rule-based approach~\cite{kim2024automatic}. 
In total, the dataset comprises 28\,010 jeonggans across 85 pieces. When counting each instrument part independently, the combined total amounts to 141\,820 jeonggans. 
Out of 90 pieces notated in jeongganbo for ensembles of at least two different instruments in the published manuscripts, we excluded 5 pieces that have discrepancies in the total number of jeonggans across instruments.

\section{Jeonggan-like Encoding}

In the field of symbolic music generation for Western monophonic and polyphonic music, encoding schemes such as ABC notation, which denotes pitch and duration separately, are effective and prevalent\cite{sturm2015folk, sturm2016music}.
However, when it comes to Korean court music, whose heterophonic structure is a defining characteristic, it is crucial that the intricate alignment of different melodies be well-represented in encoding. The genre also exhibits prolonged notes and considerable variations in note lengths, which proves to be a challenge for learning algorithms, especially when data is limited.

These distinct musical qualities call for a specialized encoding scheme; for this, we propose \textit{Jeonggan (JG)-like encoding}, which closely follows the positional notation of \textit{Jeongganbo}. This symbolic music encoding method is modeled to inherently reflect the composition and notation style of traditional Korean court music.

% (방식)
The detailed rules of encoding are as follows. The boundary of a \textit{Jeonggan} is designated as a bar (\verb`|`) token. 
Change of measure (called \textit{Gak}) is indicated by a line break (\verb!\n!). 
As illustrated in Figure~\ref{fig:encoding-position}, the position of each note is denoted by a number between 0 and 15, after which the pitch symbol follows.

Ornamentations (\textit{sigimsae}) can either have a duration or not. \textit{Sigimsae} with duration, such as the `ㄱ' symbol in Figure~\ref{fig:encoding}, are handled in the same way as pitch symbols. \textit{Sigimsae} without duration such as ` \textasciicircum{} ', which appear at the side of the pitch character, are placed after the corresponding pitch symbol.

There are several advantages that we can expect to gain from using \textit{JG-like} encoding. First, with position-based encoding, the duration-related vocabulary is limited to just 16 entries. In contrast, duration-based encoding schemes require learning each duration token as a separate entry, resulting in a significantly larger vocabulary. 
% Comparing the two cases, the Jeonggan-like encoding can be much more advantageous for learning rhythmic pattern when training with a limited dataset.
% The position-based approach has low musically grammatical problem of ensuring that the total sum of note lengths within a measure is correct for a as illustrated in Figure~\ref{fig:encoding}, score.
Additionally, rather than determining the length of a note with a single calculation, JG-like encoding allows for the flexible adjustment of note lengths during inference via combination of \textit{jeonggan} boundary and position tokens. This enables generation of music that is more adaptable to the time step and takes into account the sequence of the input source, which can be expected to result in more dynamic and context-aware music generation.

\subsection{Other Possible Encodings}
% combines bar events indicating the start of a bar and position events within a beat, thereby informing the model about the hierarchical structure between beats and bars. 
REMI (revamped MIDI-derived events)~\cite{huang2020pop} first proposed the usage of beat-position feature rather than time-shifting to encode temporal position. We also experiment with REMI-like encoding which adopts three token types: beat position, new beat (instead of new measure), and pitch tokens. 
We intentionally design REMI-like and JG-like encoding to share the same structure and result in the same number of tokens for a given melody. They differ in that JG encoding provides intra-JG position, while REMI encoding provides the beat position of the note. 
According to the position labels shown in Figure~\ref{fig:encoding-position}, any of [0, 1, 4, 10, 12] can correspond to beat position 0. However, in JG-like encoding, each occurrence of position tokens limits the possibilities of subsequent ones. For instance, a position token of 0 implies that no more notes will occur in the same \textit{jeonggan}, and if the first note is 1, one or more additional notes should follow with values of 2-3 or 6-9. In contrast, in REMI-like encoding, any offset value can follow a beat position of 0. To examine the impact of this position-based logic on the generation process, we use REMI-like encoding as our first baseline for comparison.

As a second baseline, we implement an ABC-like encoding scheme that does not have a separate bar token and encodes each note as a combination of pitch and duration values. Note that we do not omit duration tokens that are equal to unit length as ABC encoding typically does.

\section{Orchestral Part Generation}
% image attach.

\subsection{Transformer Sequence-to-sequence Model}
We implement an encoder-decoder transformer~\cite{vaswani2017attention} model to generate melodies for different instruments based on a given instrument's melody, leveraging its ability to learn long-term dependencies. Unlike RNN-based models, the transformer calculates relationships between all elements in a sequence via the self-attention mechanism, enhancing its capability in symbolic music generation\cite{huang2018music, yu2022museformer, shih2022theme}. The model consists of an encoder that processes the input sequence and a decoder for generating the output sequence. Our objective is to generate melodies that synchronize with the input melody across musically equivalent phrases; self- and cross-attention within the model enable understanding of musical context at measure and bar levels, capturing the repeating structure of melodies and accents prominent in traditional Korean court music.

\begin{figure}[!t]
\centering
  \includegraphics[alt={This image shows the model architecture used for orchestral part generation. The model encodes 1 to 5 input melodies as Note features, which are composed of tokens and instruments, and beat counters, which include previous positions, jeonggan positions, and gak positions. This encoded information is then processed by the transformer encoder and decoder to generate new melodies for the target part.},width=0.9\columnwidth]{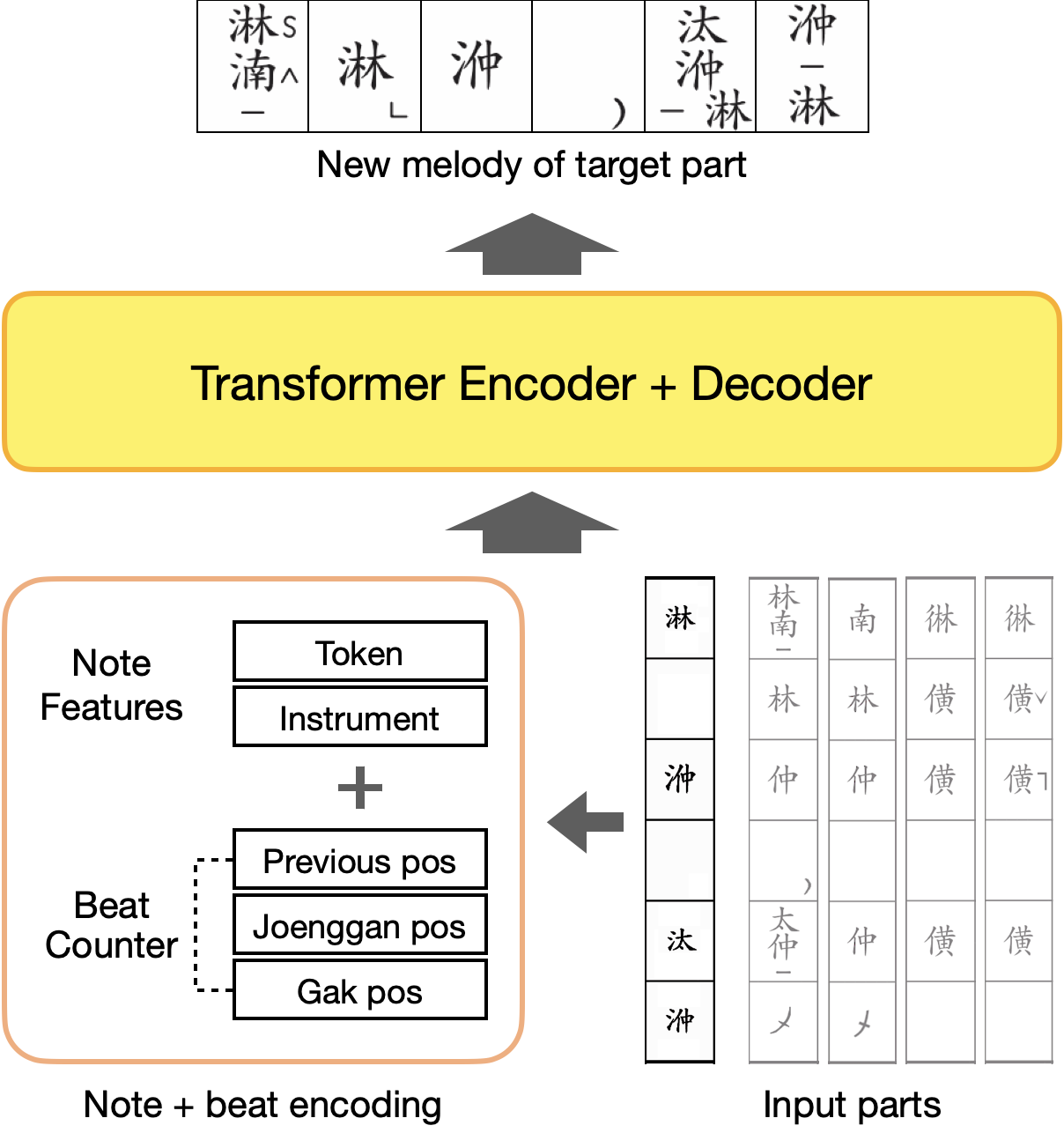}
  \caption{Orchestral part generation}
  \label{fig:Seq2seq model}
  % \vspace{-3mm}
\end{figure}

\subsection{Beat Counter}
Instead of sinusoidal~\cite{vaswani2017attention} or learned~\cite{gehring2017convolutional} positional embedding commonly utilized in transformer-based models, we implemented a `beat counter' embedding that provides information about temporal position.

For a model to learn to `parse' semantic position only from the tokens' sequential position is challenging, if not impossible, with limited training data and a small number of transformer layers. Therefore, we explicitly encode the musical position of each symbol as a combination of measure index, beat index, and sub-beat index (in-\textit{jeonggan} position) as shown in Figure~\ref{fig:Seq2seq model}. This information is summed into note embedding, just like positional encoding of transformer \cite{vaswani2017attention}. 

As previous research of PopMAG~\cite{ren2020popmag} demonstrated, metrical position embeddings can replace the positional encoding of transformers in symbolic music. A minor difference between PopMAG and our approach is that the model predicts only the appearance of new measures or new beats without the index of them, and that the new beat can be used for elongating the duration of previous note.

The same idea of embedding the beat counting has been previously applied to RNN-based Irish melody generation in ABC format~\cite{jeong2023virtuosotune}, while its advantage was not properly evaluated. A similar idea, using metrical position instead of or along with absolute token position, has also been applied to transformer architecture~\cite{ren2020popmag,  wang2021musebert, guo2023domain}. 
However, our results presented in Section~\ref{subsec:results} demonstrate that this beat counter embedding is essential for making the model properly understand the musical contents.

% 이미지 첨부하기!
% note feature: [Token, previous position, Jeonggan position, Gak position, instrument]
\section{Experiment and Results}
\subsection{Training}
We split the \textit{Jeongganbo} dataset into three subsets: 75 pieces for training, 5 for validation, and 5 for testing. Each piece contains melodies for up to 6 instruments. The sequence-to-sequence model takes 4 measures of melody, each from a randomly selected number of instruments, as input to the encoder; and given a target instrument condition, it generates the corresponding 4 measures of melody for the target instrument. Note that the number of beats in a single measure is at least 4 or to a maximum of 20 in our dataset. 

The transformer encoder and decoder both consist of 6 layers, 4 attention heads, and a hidden dimension size of 128 with dropout of 0.2. We train for 35\,000 updates across 300 epochs using negative log-likelihood loss. We also employ mixed precision training~\cite{micikevicius2017mixed} to enhance performance and efficiency. 
We utilize the Adam optimizer with an initial learning rate of 0.001 and apply a cosine learning rate scheduler with 1000 warmup steps. Using a batch size of 16, training can be conducted on a single Nvidia RTX A6000 GPU.

\subsection{Evaluation Metrics} 
\subsubsection{Length Match Rate}
As an evaluation metric, we check whether the input and output melodies share the same number of measures, a consistency necessitated by our task. Since the length of a measure can change in the middle of a piece, this metric serves as an indicator of the model's ability to capture the musical context of the input melody and accordingly generate a musically complete melody.
We measure \textit{length match rate} as the percentage of generated melodies whose number of \textit{jeonggans}, after decoding the output tokens, matches that of the input melody.

\subsubsection{F\textsubscript{1}-Score}
Regarding the generation task as one with a fixed answer, we can measure the accuracy of the generated melody by directly comparing it with the ground-truth target melody. Thus, as a general accuracy metric, we calculate the F\textsubscript{1}-score of predicted notes, where only the notes with the exact same onset position and pitch are counted as correct. To make a fair comparison between encoding methods, note onset positions in JG-like encoding were converted to those in the REMI-like format. Ornamentations without duration were not counted.

\subsection{Results and Discussion}\label{subsec:results}

\renewcommand{\arraystretch}{1.05}
\begin{table}[!t]
\centering
\small % Makes the font size smaller
\setlength{\tabcolsep}{4pt} % Adjusts the space between columns
\begin{tabular}{lcc|cc}
\toprule[1.2pt]
 & \multicolumn{2}{c|}{\textit{Piri} to \textit{Geom.}} & \multicolumn{2}{c}{Every to \textit{Daeg.}} \\
 & len-mat & F1 & len-mat & F1 \\
\midrule
JG-like & 0.942 & 0.679 & 1.0 & 0.614 \\
REMI-like & 0.923 & 0.567 & 1.0 & 0.532 \\
ABC-like & 1.0 & 0.704 & 0.903 & 0.542 \\ \hline
JG w/o Counter & 0.135 & 0.043& 0.269 & 0.052 \\
REMI w/o Counter & 0.269 & 0.081  &  0.192 & 0.039\\
ABC w/o Counter & 0.403 & 0.090 & 0.115 & 0.016 \\
\bottomrule[1.2pt]
\end{tabular}
\caption{Quantitative evaluation results}
\label{evaluation result}
% \vspace{-4mm} % Adjusts the space after the table
\end{table}

In our sequential generation process, melody for the instrument geomungo, characterized by its low pitch range and simple melodies, is the first to be generated from the initial piri melody. The daegeum, typically featuring the most complex and nuanced melodies among the six instruments, is the last in line.
Table~\ref{evaluation result} displays the results of objective evaluation, specifically focusing on geomungo and daegeum.

For generation of geomungo melodies, ABC-like encoding yields the best results. This appears to be due to the simple and regular melodic structure of the geomungo which fits in well with ABC-like encoding.
On the other hand, in the task of generating daegeum melodies, JG-like encoding achieves higher F\textsubscript{1}-scores. This indicates that JG-like encoding outperforms other methods in generating complex and varied melodies. We also discover that as rhythmic complexity increases, the measure length match rate of ABC-like encoding decreases.
% 피리-거문고에서 ABC-like model의 결과가 좋은 것은 거문고 선율이 워낙 간단하고 규칙적이기 때문에 그런 것으로 유추된다.
% 반면 대금을 생성하는 task에서는 JG encoding이 F1 score가 높은데, 대금은 음표와 시김새가 많고 리듬꼴도 다양하다. 이 결과를 통해서 즉, JG encoding이 복잡하고 다양한 멜로디를 생성할 때 다른 인코딩에 비해 유의미하게 좋은 결과를 도출함을 알 수 있다.

% \subsection{Comparison to Without Beat Counter}
To examine the effectiveness of the beat counter technique, we compare our model that incorporates beat counter with a baseline model that instead employs absolute position embedding~\cite{gehring2017convolutional}, a technique commonly used in symbolic music generation.

The results in the lower part of Table~\ref{evaluation result} show that the models without beat counter fail to generate melodies with appropriate lengths. The problem is less severe in ABC-like encoding, as processing accumulating duration tokens can be easier than counting \textit{jeonggan} boundaries. This demonstrates the efficacy of the beat counter technique in JG-like encoding, and its ability to replace traditional positional encoding.

% \section{Transforming Chwipunghyeong}
\section{15th Century Melody Transformation}
To generate an entire ensemble score using our method, we require an initial input melody with a specified instrument. However, the remaining 15th-century score of \textit{Chihwapyeong} and \textit{Chwipunghyeong} only provide a single melody without any mention of instruments. It also features rhythmic groupings of eight beats, which is rare in court music that is played today. We therefore need to transform the old melody for a specific instrument used in court music; to maintain the outline of the original melody while achieving plausible transformation, we train a masked language model on our \textit{Jeongganbo} dataset before infilling the 15th-century melody.

\subsection{BERT-like Masked Language Model}
Bidirectional Encoder Representations from Transformers (BERT)~\cite{devlin2018bert} is a self-supervised language representation learning model that uses a bidirectional transformer instead of a causal transformer decoder. It is trained with a masked language model (MLM) objective, where tokens in the input sentence are randomly masked and the model predicts the original vocabulary ID of said masked tokens. Because of its advantage in exploiting bidirectional context, BERT-like models have also been adapted for music audio generation \cite{garcia2023vampnet} and symbolic music generation \cite{dahale2022generating, casini2024investigating} along with representation-learning purpose adaptation on symbolic music \cite{wang2021musebert, zeng2021musicbert}.

\subsubsection{Piano-roll-like Encoding}
One of the main limitations of using a BERT-like model for generative tasks is that the sequence of given (unmasked) tokens and masked tokens has to be pre-defined. This means that one has to decide the number and position of new tokens to be inserted for a given original sequence. To avoid this, we use piano-roll-like encoding for the MLM, a technique widely employed in works on music generation with limited rhythmic patterns such as in Bach Chorales \cite{liang2017automatic, hadjeres2017deepbach, huang2019counterpoint, choi2023teaching}.
Here, each \textit{jeonggan} is represented as six frames, with each frame including features for symbol (pitch or \textit{sigimsae} with duration) and for ornamentation. We also apply the aforementioned beat counter in piano-roll encoding.

\subsubsection{Training with Masking}
Following examples in MusicBERT \cite{zeng2021musicbert}, we train the model with masked language model objective with various masking methods: i) masking 5\% of frames, ii) replacing 5\% of frames, iii) masking 20\% of note onsets, iv) replacing 10\% of note onsets, v) erasing 10\% of note onsets, vi) masking the entire 6 frames of 15\% of \textit{jeonggans}, and vii) masking 50\% of ornamentations.

Though the model can be trained to handle an arbitrary number of input instruments, we only train the model with a single instrument as with our orchestration transformer, since the main intended usage of the model is to create variations of a single melody. We train a 12-layer model with the same dataset and hyperparameter settings as with the orchestration model.

\subsection{Inference Procedure}
For converting and performing monophonic melodies, we opt for a 30x $\eighthnote$ span which equals to 10 \textit{jeonggans}.
% \footnote{Note that the use of `30/8' as a time signature in fig.\ref{fig:jeonggan_example} is for convenience and to avoid questions over the presence or absence of metrical levels between the  3 $\eighthnote$ beat and the 30 $\eighthnote$ phrase.}
This also corresponds to the rhythmic pattern of the 4—7th movement of \textit{Yeominlak}. The original \textit{Chihwapyeong} and \textit{Chwipunghyeong} melody, which can be interpreted in an 8/8 time signature, were modified by strategically inserting empty \textit{jeonggans} to the 5th and 7th positions, to imitate \textit{Yeominlak}'s rhythmic pattern. 
Utilizing the masked language model, the modified melodies were seamlessly transformed into a piri melody. Piri, a double-reed instrument known for its loud volume, was chosen as the main instrument for conveying the original melody due to its prominent role in contemporary court music. 

As the models were all trained on 4-measure chunks, we generate the full sequence of 512 or 132 measures using a moving window, providing two measures of previously generated output as teacher-forcing inputs and generating one more measure for each four-measure input. These were applied in a similar manner to both melody transformation and orchestral part generation.
Once the melody is transformed into a piri melody, we feed it to the orchestral transformer to generate parts for five other instruments. We sequentially generate for each instrument with the previously generated part as input. The final generation order is as follows: piri, geomungo, gayageum, ajaeng, haegeum, and daegeum.

Following the initial generation of melodies for all six instruments, we perform a refinement step. Here, each instrument's melody is regenerated with the melodies of the other five as input. This additional process helps to reinforce the melodies that initially had to be generated without the context of the other instruments.

\subsection{Expert Reviews}
% 이 부분은 비교보다는 완성 작품에 대한 평가내용
The Court Music Orchestra of the National Gugak Center performed the generated \textit{Chihwapyeong} and \textit{Chwipunghyeong} on the birth anniversary of King Sejong at Gyeongbokgung Palace on May 14th, 2024. They performed it again at the National Gugak Center on June 2nd, 2024 with an introduction to technical background by the authors. Due to time constraints, only partial excerpts from the entire score were performed.

% The generated musical pieces were presented to the performers who played them to gather their feedback.
The musicians gave positive opinions such as \textit{``genre-specific rhythm and melodic flow were well-represented''} and \textit{``the generated pieces presented ornamentation techniques and melodic progressions specialized for each instrument.''} 
% Additionally, the pieces were positively evaluated for being directly learned and generated from Jeongganbo without undergoing a transformation process to staff notation.
Still, there were a few instances where notes that did not fit the scale appeared, and when notes outside the appropriate range were present, the performers had to alter or omit them or change their octave to perform the piece. 
However, the generated results were acknowledged to closely resemble the target style of Yeominlak. Thus, the Court Music Orchestra decided to play the pieces in a similar ensemble size to Yeominlak without further modification.

% Therefore, it was suggested that additional improvements are necessary for the results to be directly used in performance without any modification.

We additionally evaluate the generated scores, focusing on the effects of the refinement step. The evaluation criteria were carefully selected to assess aspects that require a deep understanding of the genre. These criteria include 1) the appropriateness of the scale and range for each instrument (\textit{scale}), 2) the proper use of unique characteristics and ornamentations specific to each instrument (\textit{sigimsae}), 3) the suitability of the rhythmic structure of strong and weak beats (\textit{rhythm}), and 4) the harmony and coherence among the instruments when performed together as an ensemble (\textit{harmony}).

Seven employees from the National Gugak Center who majored in Korean traditional music instruments or theory participated in a subjective survey. 
We name the pre-refinement generation results piece A, and the final output after the refinement step piece B. The evaluators were not informed of this distinction.
The participants assessed the pieces for the four criteria on a 5-point scale (1-5) and provided qualitative feedback on the two compositions. The results are summarized in Table~\ref{expert evaluation}.
These results demonstrate that the proposed refinement process effectively enhances the overall quality of the generated music, especially for \textit{sigimsae} of each instrument.

\begin{table}[!t]
\centering
\begin{tabular}{lcc}\bottomrule[1.2pt]
         & \textit{No} Refinement & \textit{With} Refinement \\ \hline
Scale    & 4.0 ($\pm$0.53)         & 4.0 ($\pm$0.53)        \\
Sigimsae & 3.4 ($\pm$0.73)              & 4.0 ($\pm$0.53)        \\
Rhythm   & 2.9 ($\pm$0.35)              & 3.3 ($\pm$0.45)      \\
Harmony  & 2.9 ($\pm$0.83)              & 3.3 ($\pm$0.70)     \\ \toprule[1.2pt]
\end{tabular}
\caption{the average and std of opinion scores from 7 judges for systems with and without refinement.}
% \vspace{-2mm}
\label{expert evaluation}
\end{table}

\section{Conclusion}
% TODO
Throughout this work, we explored how music generation models can resurrect ancient melodies into new compositions that meet style of current-day Korean court music.

Venturing into relatively uncharted territory, we approached each step meticulously—from data curation and parsing to model architecture design—while carefully considering the unique nuances of the musical tradition. To enhance the quality of the generated outputs, we proposed a novel encoding framework and validated its effectiveness through objective and subjective measures. This endeavour to tackle an underrepresented non-Western music genre through diverse MIR lenses hopefully expands the horizons of the field.

The \textit{Jeongganbo} dataset and its conversion to Western staff notation in MusicXML is available online, along with other code of this project, and video recording of the performance.\footnote{\url{https://github.com/MALerLab/SejongMusic}} To the best of our knowledge, this will be the first dataset of machine-readable \textit{Jeongganbo}. We believe that this dataset can significantly contribute to computational ethnomusicology beyond its usage as a training dataset for music generation demonstrated in this paper.

We also provide an interactive web demo\footnote{\url{https://six-dragons-fly-again.site/}} that showcases our proposed generative model. While this project focused on reviving melodies from the 15th-century, the web demo allows users to input their own melodies and create orchestrations of Korean court music. 
The interactive platform enables users to directly engage with the generative model in the web browser.
% We hope the power of machine learning can make traditional music more accessible and enjoyable for modern audiences. 

% We hope that this project demonstrates the possibility of leveraging machine learning to make traditional music more accessible and enjoyable for modern audiences. 

We hope that this project contributes to moving closer to leveraging machine learning to make traditional music more accessible and enjoyable for modern audiences.

\pagebreak

\section{Acknowledgements}
We sincerely appreciate the National Gugak Center and its staff who supported this project, including director-general Kim Youngwoon (김영운), head of the research bureau Kim Myung-suk (김명석), research officers Park Jeonggyeong (박정경) and Han Jungwon (한정원). We are deeply grateful to the musicians of the Court Music Orchestra for their invaluable contributions and efforts to vitalize our humble results.
This research was also supported by the National R\&D Program through the National Research Foundation of Korea (NRF) funded by the Korean Government (MSIT) (RS-2023-00252944, Korean Traditional Gagok Generation Using Deep Learning).

% For bibtex users:
\bibliography{ISMIR2024template}

% For non bibtex users:
%\begin{thebibliography}{citations}
% \bibitem{Author:17}
% E.~Author and B.~Authour, ``The title of the conference paper,'' in {\em Proc.
% of the Int. Society for Music Information Retrieval Conf.}, (Suzhou, China),
% pp.~111--117, 2017.
%
% \bibitem{Someone:10}
% A.~Someone, B.~Someone, and C.~Someone, ``The title of the journal paper,''
%  {\em Journal of New Music Research}, vol.~A, pp.~111--222, September 2010.
%
% \bibitem{Person:20}
% O.~Person, {\em Title of the Book}.
% \newblock Montr\'{e}al, Canada: McGill-Queen's University Press, 2021.
%
% \bibitem{Person:09}
% F.~Person and S.~Person, ``Title of a chapter this book,'' in {\em A Book
% Containing Delightful Chapters} (A.~G. Editor, ed.), pp.~58--102, Tokyo,
% Japan: The Publisher, 2009.
%
%
%\end{thebibliography}

\end{document}